\documentclass[a4paper,12pt,eqno]{article}
\usepackage{theorem}
\usepackage{latexsym,amssymb,amsfonts,amsmath}
\newtheorem{thm}{Theorem}[section]
\newtheorem{lem}[thm]{Lemma}
\newtheorem{cor}[thm]{Corollary}
\newtheorem{pro}[thm]{Proposition}

\theoremheaderfont{\scshape}
\newcommand{\RM}{\mathbb{R}}
\newcommand{\ZM}{\mathbb{Z}}

\newcommand{\CM}{\mathbb{C}}

\newcommand{\PM}{\mathbb{P}}

\newcommand{\ket}[1]{|#1\rangle}

\title{{\Large {\bf One-dimensional discrete-time quantum walks \\ 
on random environments}}}
\author{
{\small Norio Konno}\\
{\scriptsize Department of Applied Mathematics, 
Faculty of Engineering, 
Yokohama National University}\\
{\scriptsize Hodogaya, Yokohama 240-8501, Japan}\\
{\scriptsize e-mail: konno@ynu.ac.jp}\\
}
\date{\empty }
\begin{document}
\maketitle

\par\noindent
\begin{small}
\par\noindent
{\bf Abstract}. We consider discrete-time nearest-neighbor quantum walks on random environments in one dimension. Using the method based on a path counting, we present both quenched and annealed weak limit theorems for the quantum walk. 
\footnote[0]{
{\it Abbr. title:} Quantum walks on random environments
}
\end{small}

\setcounter{equation}{0}
\section{Introduction}
Classical random walks have proved to be a very important tool for various fields. Quantum walks are the quantum counterpart of classical random walks. There are two types of quantum walks. One is the discrete-time walk and the other is the continuous-time one. Reviews and books on quantum walks are Kempe \cite{Kempe2003}, Kendon \cite{Kendon2007}, Venegas-Andraca \cite{VAndraca2008}, Konno \cite{Konno2008a, Konno2008b}, for examples. Ambainis {\it et al.} \cite{AmbainisEtAl2001} intensively studied discrete-time quantum walks on $\ZM$, where $\ZM$ is the set of integers. Here we consider a discrete-time nearest-neighbor quantum walk in a random environment (QWRE) on $\ZM$. We present weak limit theorems for the QWRE, both quenched (i.e., conditional upon the environment) and annealed (i.e., averaged over the environment), by counting the number of paths that takes a quantum walker from the origin to a position. Numerical results on continuous-time QWRE were reported in Yin et al. \cite{Yin2008}. As for the classical case, the study of random walks in random environments was initiated in the mid-1970s and has been investigated by many researchers. Sinai's random walk in a random environment \cite{Sinai1982} is one of the typical models. See Zeitouni \cite{Zeitouni2004} for a review of random walks on random environments. To the best of our knowledge, any rigorous result for the QWRE is not known. Therefore we hope that our investigation would be a first step towards the study of the QWRE. 

The rest of the paper is organized as follows. Section 2 gives the definition of the QWRE. In Sect. 3, we present the quenched weak limit theorem (Theorem {\rmfamily \ref{thm2}}). As a consequence, the annealed weak limit theorem (Corollary {\rmfamily \ref{corA}}) is also obtained. Section 4 is devoted to the proof of Theorem {\rmfamily \ref{thm2}}.

\section{Definition}
The time evolution of the QWRE on $\ZM$ is determined by a sequence of $2 \times 2$ random unitary matrices, $\{ U_x : x \in \ZM \}$, where
\begin{eqnarray*}
U_x =
\left[
\begin{array}{cc}
a_x & b_x \\
c_x & d_x
\end{array}
\right],
\end{eqnarray*}
with $a_x,b_x,c_x,d_x \in \mathbb C$ and $\CM$ is the set of complex numbers.  The subscript $x$ indicates the location. The unitarity of $U_x$ gives 
\begin{eqnarray*}
|a_x|^2 + |c_x|^2 =|b_x|^2 + |d_x|^2 =1, \> a_x \overline{c_x} + b_x \overline{d_x}=0, \> c_x= - \triangle_x \overline{b_x}, \> d_x= \triangle_x \overline{a_x},\label{konno-eqn:seisitu}
\end{eqnarray*}
where $\overline{z}$ is the complex conjugate of $z \in \mathbb C$ and $\triangle_x = \det U_x = a_x d_x - b_x c_x$ with $|\triangle_x|=1.$

The discrete-time quantum walk is a quantum version of the classical random walk with additional degree of freedom called chirality. We should remark that there is a strong structural similarity between quantum walks and correlated random walks, see Konno \cite{Konno2009}. The chirality takes values left and right, and it means the direction of the motion of the particle. At each time step, if the particle has the left chirality, it moves one step to the left, and if it has the right chirality, it moves one step to the right. Let define
\begin{eqnarray*}
\ket{L} = 
\left[
\begin{array}{cc}
1 \\
0  
\end{array}
\right],
\qquad
\ket{R} = 
\left[
\begin{array}{cc}
0 \\
1  
\end{array}
\right], 
\end{eqnarray*}
so $U_x$ acts on two chiralities as follows: 
\begin{eqnarray*}
U_x \ket{L} = a_x \ket{L} + c_x \ket{R}, \qquad U_x \ket{R} = b_x \ket{L} + d_x \ket{R},  
\end{eqnarray*}
where $L$ and $R$ refer to the left and right chirality state, respectively.  To define the dynamics of the QWRE, we divide $U_x$ into two matrices:
\begin{eqnarray*}
P_x =
\left[
\begin{array}{cc}
a_x & b_x \\
0 & 0 
\end{array}
\right], 
\quad
Q_x=
\left[
\begin{array}{cc}
0 & 0 \\
c_x & d_x 
\end{array}
\right],
\end{eqnarray*}
with $U_x=P_x+Q_x$. The important point is that $P_x$ (resp. $Q_x$) represents that the particle moves to the left (resp. right) at position $x$ at each time step.

We will explain the QWRE more precisely. Let $\Omega = \RM^{\ZM}$, where $\RM$ is the set of real numbers. A random environment is an $\Omega$-valued random variable $\omega = \{ \omega_x : x \in \ZM \}$ with probability measure $\bar{P}$. We will assume that $\bar{P}$ is a product measure on $(\Omega, {\cal F})$, where ${\cal F}$ is the Borel $\sigma$-field of $\Omega$. So we write $\bar{P}= \prod_{x \in \ZM} \bar{P}_x$, where $\bar{P}_x$ is a probability measure on $(\RM, {\cal G})$ and ${\cal G}$ is the Borel $\sigma$-field of $\RM$. Throughout this paper, we consider the following $U_x$: 
\begin{eqnarray*}
U_x =
\frac{1}{\sqrt{2}}
\left[
\begin{array}{cc}
e^{i \omega_x} & 1 \\
1 & -e^{-i \omega_x} 
\end{array}
\right].
\end{eqnarray*}
A typical example is that $\omega = \{ \omega_x : x \in \ZM\}$ is independent and identically distributed. That is, $\bar{P}_x$ does not depend on position $x$. In particular, when $\omega = {\bf 0}$, i.e., $\omega_x = 0 \> (x \in \ZM)$ or $\bar{P}_x = \delta_0$, where $\delta_a$ is the delta measure at $a \in \RM$, the dynamics is non-random and position-independent. This case is equivalent to the {\it Hadamard walk} determined by the Hadamard gate $U_x \equiv H$:
\begin{eqnarray*}
H=\frac{1}{\sqrt2}
\left[
\begin{array}{cc}
1 & 1 \\
1 &-1 
\end{array}
\right].
\end{eqnarray*}
The walk has been extensively investigated in the study of the quantum walk.

The set of initial qubit states at the origin for the QWRE is given by 
\begin{eqnarray*}
\Phi = \left\{ \varphi =
{}^T [\alpha,\beta] \in \mathbb C^2 :
|\alpha|^2 + |\beta|^2 =1
\right\},
\end{eqnarray*}
where $T$ is the transposed operator. In the present paper, we take $\varphi_{\ast} = {}^T [1/\sqrt{2},i/\sqrt{2}]$ as the initial qubit state of the QWRE. Then the probability distribution of the Hadamard walk starting from $\varphi_{\ast}$ at the origin is symmetric. We call the quantum walk the {\it symmetric} Hadamard walk here.

Let $\Xi_{n}(l, m)$ denote the sum of all paths starting from the origin in the trajectory consisting of $l$ steps left and $m$ steps right. In fact, for time $n = l+m$ and position $x=-l + m$, we have 
\begin{align*}
\Xi_n (l,m) = \sum_{l_j, m_j} P^{l_{1}}_{x(l_{1})} Q^{m_{1}}_{x(m_{1})} P^{l_{2}}_{x(l_{2})} Q^{m_{2}}_{x(m_{2})} \dots P^{l_{n-1}}_{x(l_{n-1})} Q^{m_{n-1}}_{x(m_{n-1})} P^{l_{n}}_{0} Q^{m_{n}}_{0},
\end{align*}
where the summation is taken over all $l_j, m_j \ge 0$ satisfying $l_1+ \cdots +l_n=l, \> m_1+ \cdots + m_n = m, \> l_j+ m_j=1,$ and $|x(a_{j+1})- x(b_{j})| = 1, \> x(a_{j+1}), \> x(b_{j}) \in \ZM \> (a, b \in \{l,m\})$. We should note that the definition gives 
\begin{align*}
\Xi_{n+1}(l, m) = P_{x+1} \> \Xi_{n}(l-1, m) + Q_{x-1} \> \Xi_{n}(l, m-1).
\end{align*}

For example, in the case of $l=3, \> m=1$, we have 
\begin{align}
\Xi_4 (3,1) &= Q_{-3} P_{-2} P_{-1} P_0 + P_{-1} Q_{-2} P_{-1} P_0 + P_{-1} P_0 Q_{-1} P_0 + P_{-1} P_0 P_1 Q_0. 
\label{kaede}
\end{align}

The probability that a quantum walker in an environment $\omega \in \Omega$ is in position $x$ at time $n$ starting from the origin with $\varphi_{\ast} (={}^T [1/\sqrt{2},i/\sqrt{2}]) \in \Phi$ is defined by 
\begin{align*}
P^{\omega}_n (X_{n} =x) := || \Xi_{n}(l, m) \varphi_{\ast} ||^2,
\end{align*}
where $n=l+m$ and $x=-l+m$.

For each $\omega \in \Omega$, $P^{\omega}_n$ is a probability measure on $(\ZM, {\cal H})$, where ${\cal H}$ is the set of all subsets of $\ZM$. Thus we can define a probability measure $\PM_n := \bar{P} \otimes P_n ^{\omega}$ on $\Omega \times \ZM$ by the formula 
\begin{align*}
\PM_n (H \times \{ X_n \in A \}) := \int_H \> P_n^{\omega} (\{ X_n \in A \}) \> \bar{P} (d \omega),
\end{align*}
for any $H \in {\cal H}, \> A \subset \ZM$. Statements involving $P_n ^{\omega}$ are called $quenched$. On the other hand, statements involving $\PM_n$ are called $annealed$. Expectations under $\bar{P}$ and $P_n ^{\omega}$ are denoted by $\bar{E}$ and $E_n ^{\omega}$, respectively. In $n=4$ case, by definition, a direct computation implies
\begin{align*}
P^{\omega} _4 (X_4 = -4) 
&= \frac{1 + \sin (\omega_0)}{2^4}, \quad
P^{\omega} _4 (X_4 = -2) 
= \frac{6 + 4 \sin (\omega_0)}{2^4}, \\
P^{\omega} _4 (X_4 = 0) 
&= \frac{2}{2^4}, \quad
P^{\omega} _4 (X_4 = 2) 
= \frac{6 - 4 \sin (\omega_0)}{2^4}, \quad
P^{\omega} _4 (X_4 = 4)  
= \frac{1 - \sin (\omega_0)}{2^4}.
\end{align*}
If the probability measure of $\omega_0$, i.e., $\bar{P}_0$, is symmetric, (for example, $\omega_0$ is a uniform distribution on $[- \pi, \pi]$), then 
\begin{align*}
\PM_4 (X_4 = -4) 
&= \frac{1}{2^4}, \quad
\PM_4 (X_4 = -2) 
= \frac{6}{2^4}, \quad 
\PM_4 (X_4 = 0) 
= \frac{2}{2^4}, \\
\PM_4 (X_4 = 2) 
&= \frac{6}{2^4}, \quad
\PM_4 (X_4 = 4)  
= \frac{1}{2^4}.
\end{align*}
In fact, Corollary {\rmfamily \ref{corenaji}} tells us that $\PM_n (X_n = \cdot)$ is the same as the probability distribution of the symmetric Hadamard walk for any $n \ge 0$.  

\section{Results}
In this section, we will present both quenched and annealed weak limit theorems for the QWRE. Quantum walks behave quite differently from classical random walks. For example, in the classical case, the probability distribution is a binomial one. On the other hand, the probability distribution of the Hadamard walk has a complicated and oscillatory form. For the classical case, the well-known central limit theorem holds. That is, $Y_n/\sqrt{n}$ converges to a normal distribution as $n \to \infty$, where $Y_n$ denotes the probability distribution of the location for the random walker at time $n$. Concerning the quantum walk defined by $2 \times 2$ non-random unitary matrix $U_x \equiv U$, Konno \cite{Konno2002, Konno2005} showed the corresponding weak limit theorem. In particular, for the symmetric Hadamard walk, i.e., $\omega \equiv {\bf 0}$, we have  
\begin{thm}
\label{thm1}
\begin{align*}
\lim_{n \to \infty} P_n ^{\bf 0} \left( u \leq \frac{X_{n}}{n} \leq v \right) 
= \int_{u}^{v} \> f_K (x) \> dx, 
\end{align*}
where $-\infty < u \le v < \infty$, and
\[
f_K (x) = \frac{1}{ \pi ( 1-x^2 ) \sqrt{1-2 x^2} } \> I_{(-1/\sqrt{2}, \> 1/\sqrt{2})}(x).
\]
Here $I_A (x) = 1 \> (x \in A), \> = 0 \> (x \not\in A).$ 
\end{thm}
We can extend this result ($\omega \equiv {\bf 0}$) to the general $\omega \in \Omega$. The following {\it quenched} weak limit theorem is our main result of this paper.  
\begin{thm}
\label{thm2}
(Quenched case for $X_n/n$). For any fixed environment $\omega \in \Omega$, 
\begin{align*}
\lim_{n \to \infty} P_n ^{\omega} \left( u \leq \frac{X_{n}}{n} \leq v \right) 
= \int_{u}^{v} \> \{ 1 - \sin (\omega_0) x \} \> f_K (x) \> dx, 
\end{align*}
where $-\infty < u \le v < \infty.$ 
\end{thm}
One of the interesting points of this result is that the limit density depends only on $\omega_0$. By Theorem {\rmfamily \ref{thm2}}, we can obtain an {\it annealed} weak limit theorem immediately as follows:
\begin{cor}
\label{corA}
(Annealed case for $X_n/n$) 
\begin{align*}
\lim_{n \to \infty} \PM_n \left( u \leq \frac{X_{n}}{n} \leq v \right) 
= \int_{u}^{v} \> \{ 1 - \bar{E}(\sin (\omega_0)) x \} \> f_K (x) \> dx, 
\end{align*}
where $- \infty < u \le v < \infty.$ In particular, when the probability measure of $\omega_0$, (i.e., $\bar{P}_0$) is symmetric, we have
\begin{align*}
\lim_{n \to \infty} \PM_n \left( u \leq \frac{X_{n}}{n} \leq v \right) 
= \int_{u}^{v} \> f_K (x) \> dx.
\end{align*}
\end{cor}
If the probability measure of $\omega_0$ is symmetric, then the limit density is the same as that of Theorem {\rmfamily \ref{thm1}} for the symmetric Hadamard walk. 

If $\omega_x = \pi/2 \> (x \in \ZM)$, then
\begin{eqnarray*}
U_x \equiv
\frac{1}{\sqrt{2}}
\left[
\begin{array}{cc}
i & 1 \\
1 & i 
\end{array}
\right].
\end{eqnarray*}
In this non-random case, when we take $\varphi_{\ast} = {}^T [1/\sqrt{2},i/\sqrt{2}]$ as the initial qubit state, the weak limit theorem shown by Konno \cite{Konno2002, Konno2005} gives the same conclusion as Theorem {\rmfamily \ref{thm2}}. That is, the limit density is $( 1 - x ) \> f_K (x).$

From Theorem {\rmfamily \ref{thm2}}, for each environment $\omega \in \Omega$, the variance of the limit density, $V^{\omega}$, is given by 
\begin{cor}
\label{corvar} 
\[
V^{\omega} = \frac{2-\sqrt{2}}{2} \> \left\{ 1 - \frac{2-\sqrt{2}}{2} \sin^2( \omega_0) \right\}.
\]
\end{cor} 
When $\omega_0 = 0$ (i.e., $\bar{P}_0 = \delta_0$), $V^{\omega}$ takes the maximum value. Therefore the additional noise makes the variance smaller. On the other hand, if $\omega_0 = \pi/2$ (i.e., $\bar{P}_0 = \delta_{\pi/2}$), then $V^{\omega}$ takes the minimum value. So the additional noise makes the variance larger.  As for random walks in random environments, for example, the following result for the Sinai random walk corresponding to a usual symmetric random walk (i.e., in a non-random environment) is known. Let $Z_n$ denote the probability distribution of the location for the random walker at time $n$ in a random environment. Then $Z_n/(\log n)^2$ converges to some non-degenerate random variable as $n \to \infty$, (see \cite{Sinai1982, Zeitouni2004} for more details). The result tells us that this order $(\log n)^2$ is smaller than the order $\sqrt{n}$ for the non-random environment case. On the other hand, our Theorems {\rmfamily \ref{thm1}} and {\rmfamily \ref{thm2}} imply that both quantum walks in non-random and random environments have the same order $n$. The difference between them is the magnitude of the variance as shown in Corollary {\rmfamily \ref{corvar}}. Moreover, compared with the non-random environment case, the variance for the corresponding random environment case can be larger or smaller. 

In our setting, $\{ U_x : x \in \ZM \}$ is a sequence of position-dependent $2 \times 2$ random unitary matrices. On the other hand, when $\{ U_n : n =1, 2, \ldots \}$ is a sequence of time-dependent $2 \times 2$ random unitary matrices, Mackay et al. \cite{MackayEtAl2002} and Ribeiro et al. \cite{RibeiroEtAl2004} investigated and showed that the probability distribution of the quantum walk converges to a binomial distribution by averaging over many trials by numerical simulations. Konno \cite{Konno2005b} proved their results by using a path counting method. In this paper, we use a similar approach to prove Theorem {\rmfamily \ref{thm2}}.

We discuss other related works \cite{BrunEtAl2003a, BrunEtAl2003b, SegawaKonno2008, Richter2007a, Richter2007b, ErmannEtAl2006}. Brun et al. \cite{BrunEtAl2003a, BrunEtAl2003b} considered discrete-time quantum walks with many coins on the line. By a different method, they analytically showed that the number of different coins has one of the same effects noted in our work: the more different coins are used, the smaller the variance, while still retaining the quantum scaling. The corresponding weak limit theorem was presented in \cite{SegawaKonno2008}. Richter \cite{Richter2007a, Richter2007b} proved that decoherent quantum walks on finite sets of positions under repeated randomized measurements have the potential speed up a large class of their classical counterparts. This work is still analytic using yet another method. Ermann et al. \cite{ErmannEtAl2006} investigated discrete-time quantum walks on the line in chaotic and regular environments by using the method given in \cite {BrunEtAl2003a, BrunEtAl2003b}. The family of chaotic environments includes the so-called quantum multibaker map as a special case. Their analytical results suggest that chaotic environments induce decoherence more effectively than regular ones.

\section{Proof of Theorem {\rmfamily \ref{thm2}}}
In this section we will give the proof of Theorem {\rmfamily \ref{thm2}}. First we introduce the useful random matrices to compute $\Xi_n (l,m)$:
\[
R_x =
\left[
\begin{array}{cc}
c_x  & d_x \\
0 & 0 
\end{array}
\right], 
\quad
S_x =
\left[
\begin{array}{cc}
0 & 0 \\
a_x & b_x 
\end{array}
\right].
\]
In general, we obtain the following table of products of the matrices $P_x,Q_x,R_x$ and $S_x$ ($x \in \ZM)$: for any $x, y \in \ZM,$ 
\par
\
\par
\begin{center}
\begin{tabular}{c|cccc}
  & $P_y$ & $Q_y$ & $R_y$ & $S_y$  \\ \hline
$P_x$ & $a_xP_y$ & $b_xR_y$ & $a_xR_y$ & $b_xP_y$  \\
$Q_x$ & $c_xS_y$ & $d_xQ_y$ & $c_xQ_y$ & $d_xS_y$ \\
$R_x$ & $c_xP_y$ & $d_xR_y$ & $c_xR_y$ & $d_xP_y$ \\
$S_x$ & $a_xS_y$ & $b_xQ_y$ & $a_xQ_y$ & $b_xS_y$ 
\end{tabular}
\end{center}
\par
\
\par\noindent
where $P_x Q_y = b_x R_y$, for example. From this table and Eq. ({\rmfamily \ref{kaede}}), we obtain
\begin{eqnarray}
\label{konno-eqn:araki}
\Xi_4 (3,1) = 
(b_{-1} c_{-2} a_{-1} + a_{-1} b_0 c_{-1} ) P_0 + a_{-1} a_0 b_1 R_0 
+ c_{-3} a_{-2} a_{-1} S_0. 
\end{eqnarray}
We should note that $P_0, Q_0, R_0$ and $S_0$ form an orthonormal basis of the vector space of complex $2 \times 2$ matrices with respect to the trace inner product $\langle A | B \rangle = $ tr$(A^{\ast}B)$, where $\ast$ means the adjoint operator. So $\Xi_n (l,m)$ has the following form:
\begin{eqnarray*}
\Xi_n (l,m) = p_n (l,m) P_0 + q_n (l,m) Q_0 + r_n (l,m) R_0 + s_n (l,m) S_0.
\label{konno-eqn:yukitan}
\end{eqnarray*}
As in a similar way for the one-dimensional discrete-time quantum walk (see Konno, \cite{Konno2002, Konno2005}, for example), we can compute explicit forms of $p_n (l,m), q_n (l,m), r_n (l,m)$ and $s_n (l,m)$ as follows: 

\begin{pro}
\label{pro1}
When $l \wedge m (:= \min \{ l, m \}) \ge 1$, we obtain 
\begin{align*}
p_n (l,m) &= \Theta^{(p)}_n (l,m) \> p^{(H)}_n (l,m), \qquad
q_n (l,m)  = \Theta^{(q)}_n (l,m) \> q^{(H)}_n (l,m), \\
r_n (l,m) &= \Theta^{(r)}_n (l,m) \> r^{(H)}_n (l,m), \qquad
s_n (l,m)  = \Theta^{(s)}_n (l,m) \> s^{(H)}_n (l,m),
\end{align*}
where 
\begin{align*}
\Theta^{(p)}_n (l,m) &= 
\begin{cases} 
\displaystyle{e^{i(\omega_{-1} + \cdots + \omega_{-(l-m-1)})}} & \text{if $l-1 > m$}, \\ 
1 & \text{if $l-1 = m$}, \\
\displaystyle{e^{-i(\omega_{0} + \cdots + \omega_{m-l})}} & \text{if $l-1 < m$}
\end{cases}
\\
\Theta^{(q)}_n (l,m) &= 
\begin{cases} 
\displaystyle{e^{i(\omega_{0} + \cdots + \omega_{-(l-m)})}} & \text{if $l > m-1$}, \\ 
1 & \text{if $l= m-1$}, \\
\displaystyle{e^{-i(\omega_{1} + \cdots + \omega_{m-l-1})}} & \text{if $l < m-1$}, 
\end{cases}
\\
\Theta^{(r)}_n (l,m) &= 
\begin{cases} 
\displaystyle{e^{i(\omega_{0} + \cdots + \omega_{-(l-m-1)})}} & \text{if $l > m$}, \\ 
1 & \text{if $l= m$}, \\
\displaystyle{e^{-i(\omega_{1} + \cdots + \omega_{m-l})}} & \text{if $l < m$}, 
\end{cases}
\\
\Theta^{(s)}_n (l,m) &= 
\begin{cases} 
\displaystyle{e^{i(\omega_{-1} + \cdots + \omega_{-(l-m)})}} & \text{if $l > m$}, \\ 
1 & \text{if $l= m$}, \\
\displaystyle{e^{-i(\omega_{0} + \cdots + \omega_{m-l-1})}} & \text{if $l < m$}, 
\end{cases}
\end{align*}
and
\begin{align*}
p^{(H)}_n (l,m) &= \left( \frac{1}{\sqrt{2}} \right)^{n-1} \> \sum_{\gamma =1}^{(l-1) \wedge m} (-1)^{m- \gamma} {l-1 \choose \gamma} {m-1 \choose \gamma -1}, \\
q^{(H)}_n (l,m) &= \left( \frac{1}{\sqrt{2}} \right)^{n-1} \> \sum_{\gamma =1}^{l \wedge (m-1)} (-1)^{m- \gamma -1} {l-1 \choose \gamma -1} {m-1 \choose \gamma}, \\
r^{(H)}_n (l,m) &= s^{(H)}_n (l,m) =\left( \frac{1}{\sqrt{2}} \right)^{n-1} \> \sum_{\gamma =1}^{l \wedge m} (-1)^{m- \gamma} {l-1 \choose \gamma -1} {m-1 \choose \gamma -1}.
\end{align*}
When $l \wedge m =0$, we have
\begin{align*}
p^{(H)}_n (0,n) 
&= r^{(H)}_n (0,n) = s^{(H)}_n (0,n) =0, \quad q^{(H)}_n (0,n) = (-1/\sqrt{2})^{n-1}, \\
\Theta^{(p)}_n (0,n) &= \Theta^{(r)}_n (0,n) = \Theta^{(s)}_n (0,n) =1, \quad \Theta^{(q)}_n (0,n) =e^{-i(\omega_1 + \cdots + \omega_{n-1})}, 
\end{align*}
and
\begin{align*}
p^{(H)}_n (n,0) 
&= (1/\sqrt{2})^{n-1}, \quad q^{(H)}_n (n,0) = r^{(H)}_n (n,0) = s^{(H)}_n (n,0) = 0, \\
\Theta^{(p)}_n (n,0) &= e^{-i(\omega_{-1} + \cdots + \omega_{-(n-1)})}, \quad \Theta^{(q)}_n (n,0) = \Theta^{(r)}_n (n,0) = \Theta^{(s)}_n (n,0) =1.  
\end{align*}
\end{pro}

We should remark that if $\omega =$ {\bf 0} (i.e., Hadamard walk), then $\Theta^{(p)}_n (l,m) = \Theta^{(q)}_n (l,m) = \Theta^{(r)}_n (l,m) = \Theta^{(s)}_n (l,m) \equiv 1.$ Therefore we use superscript $(H)$ for $a^{(H)}_n (l,m) \> (a=p,q,r,s)$. For example, in $l=3, m=1$ case, Proposition {\rmfamily \ref{pro1}} gives 
\begin{align*}
\Theta^{(p)}_4 (3,1) &= e^{i \omega_{-1}}, \quad \Theta^{(q)}_4 (3,1) = e^{i (\omega_0 + \omega_{-1} + \omega_{-2})}, \\
\Theta^{(r)}_4 (3,1) &= e^{i (\omega_0 + \omega_{-1})}, \quad \Theta^{(s)}_4 (3,1) =e^{i (\omega_{-1} + \omega_{-2})}, \\
p^{(H)}_4 (3,1) &=2(1/\sqrt{2})^3 , \quad  q^{(H)}_4 (3,1) =0, \quad  r^{(H)}_4 (3,1) = s^{(H)}_4 (3,1) =(1/\sqrt{2})^3.    
\end{align*}
On the other hand, by Eq.({\rmfamily \ref{konno-eqn:araki}}), 
\begin{align*}
\Xi_4 (3,1) = (1/\sqrt{2})^3 \left( 2 e^{i \omega_{-1}} P_0 + e^{i (\omega_0 + \omega_{-1})} R_0 + e^{i (\omega_{-1} + \omega_{-2})} S_0 \right).
\end{align*}
So we have the same conclusion. From Proposition {\rmfamily \ref{pro1}}, we have
\begin{cor}
If $n=l+m$ and $x=-l+m$, then 
\begin{align*}
P^{\omega}_n (X_n = x) 
&= P^{{\bf 0}}_n (X_n = x) + \frac{1}{2} \left\{ p^{(H)}_n (l,m)^2 - q^{(H)}_n (l,m)^2 \right\} \cdot \sin (\omega_0) \\
&- \frac{r^{(H)}_n (l,m)}{2} \left[ p^{(H)}_n (l,m) \> \Im \left( (1+e^{2i \omega_0}) \> \Theta^{(p)}_n (l,m) \> \overline{\Theta^{(r)}_n (l,m)} \right) \right. \\
&\left. \qquad \qquad \qquad +  q^{(H)}_n (l,m) \> \Im \left( (1+e^{2i \omega_0}) \> \Theta^{(s)}_n (l,m) \> \overline{\Theta^{(q)}_n (l,m)} \right) \right] \\
&=: P^{{\bf 0}}_n (X_n = x) + W_1 (\omega_0) + W_2 (\omega),
\end{align*}
where $\Im (z)$ is the imaginary part of $z \in \CM.$
\end{cor}
From now on, we will show that $W_2 (\omega)=0$ for any $l,m$. First we consider $l-1 > m$. In this case, we have
\begin{align*}
\Theta^{(p)}_n (l,m) &= e^{i(\omega_{-1} + \cdots + \omega_{-(l-m-1)})}, \quad
\Theta^{(q)}_n (l,m) = e^{i(\omega_{0} + \cdots + \omega_{-(l-m)})}, \\
\Theta^{(r)}_n (l,m) &= e^{i(\omega_{0} + \cdots + \omega_{-(l-m-1)})}, \quad 
\Theta^{(s)}_n (l,m) = e^{i(\omega_{-1} + \cdots + \omega_{-(l-m)})}.
\end{align*}
By using the above equations,
\begin{align*}
\Im \left( (1+e^{2i \omega_0}) \> \Theta^{(p)}_n (l,m) \> \overline{\Theta^{(r)}_n (l,m)} \right) 
&= \Im \left( (1+e^{2i \omega_0}) \> \Theta^{(s)}_n (l,m) \> \overline{\Theta^{(q)}_n (l,m)} \right) \\
&= \Im \left( (1+e^{2i \omega_0}) e^{-i \omega_0} \right) =0.
\end{align*}
Therefore we obtain $W_2 (\omega) =0$. Next we consider $l-1 = m$ case. Then we have
\begin{align*}
\Theta^{(p)}_n (l,m) &= 1, \quad
\Theta^{(q)}_n (l,m) = e^{i(\omega_{0} + \omega_{-1})}, \\
\Theta^{(r)}_n (l,m) &= e^{i\omega_{0}}, \quad 
\Theta^{(s)}_n (l,m) = e^{i \omega_{-1}}.
\end{align*}
These imply 
\begin{align*}
\Im \left( (1+e^{2i \omega_0}) \> \Theta^{(p)}_n (l,m) \> \overline{\Theta^{(r)}_n (l,m)} \right) 
&= \Im \left( (1+e^{2i \omega_0}) \> \Theta^{(s)}_n (l,m) \> \overline{\Theta^{(q)}_n (l,m)} \right) \\
&= \Im \left( (1+e^{2i \omega_0}) e^{-i \omega_0} \right) =0.
\end{align*}
So we get $W_2 (\omega) =0$. For other cases, we have the same conclusion in a similar fashion. As a consequence,

\begin{pro}
\label{pro2}
\begin{align*}
P^{\omega}_n (X_n = x) = P^{{\bf 0}}_n (X_n = x) + \frac{1}{2} \left\{ p^{(H)}_n (l,m)^2 - q^{(H)}_n (l,m)^2 \right\} \cdot \sin (\omega_0),
\end{align*}
where $n=l+m$ and $x=-l+m$.
\end{pro}
Then we have immediately
\begin{cor}
\label{corenaji}
If the probability measure of $\omega_0$ (i.e., $\bar{P}_0$), is symmetric, then
\begin{align*}
\PM_n (X_n = x) = P^{{\bf 0}}_n (X_n = x),
\end{align*}
for any $n \ge 0$ and $x \in \ZM$.
\end{cor}

Let $[x]$ denote the integer part of $x \in \RM$. For $1 \le l \le [n/2]$, $p^{(H)}_n (l,m)$ and $q^{(H)}_n (l,m)$ can be rewritten as   
\begin{align*}
p^{(H)}_n (l,m) 
&= \left( \frac{1}{\sqrt{2}} \right)^{n-1} \> (-1)^{n-l-1} \> l \> \sum_{\gamma =1}^{l} (-1)^{\gamma-1} \frac{1}{\gamma} {l-1 \choose \gamma-1} {n-l-1 \choose \gamma -1} \\
&+ \left( \frac{1}{\sqrt{2}} \right)^{n-1} \> (-1)^{n-l}  \> \sum_{\gamma =1}^{l} (-1)^{\gamma-1} {l-1 \choose \gamma-1} {n-l-1 \choose \gamma -1}, \\
q^{(H)}_n (l,m) 
&= \left( \frac{1}{\sqrt{2}} \right)^{n-1} \> (-1)^{n-l} \> (n-l) \> \sum_{\gamma =1}^{l} (-1)^{\gamma-1} \frac{1}{\gamma} {l-1 \choose \gamma-1} {n-l-1 \choose \gamma -1} \\
&- \left( \frac{1}{\sqrt{2}} \right)^{n-1} \> (-1)^{n-l} \> \sum_{\gamma =1}^{l} (-1)^{\gamma-1} {l-1 \choose \gamma-1} {n-l-1 \choose \gamma -1}. \\
\end{align*}
Furthermore we will rewrite $p^{(H)}_n (l,m)$ and $q^{(H)}_n (l,m)$ by using Jacobi polynomials. Let $P^{\nu, \mu} _n (x)$ denote the Jacobi polynomial which is orthogonal on $[-1,1]$ with respect to $(1-x)^{\nu}(1+x)^{\mu}$ with $\nu, \mu > -1$. Then the following relation holds:
\begin{eqnarray}
P^{\nu, \mu} _n (x) = \frac{\Gamma (n + \nu + 1)}{\Gamma (n+1) \Gamma (\nu +1)} \> {}_2F_1(- n, n + \nu + \mu +1; \nu +1 ;(1-x)/2),
\label{konno-eqn:ken}
\end{eqnarray}
where ${}_2F_1(a, b; c ;z)$ is the hypergeometric series and $\Gamma (z)$ is the gamma function. In general, as for orthogonal polynomials, see \cite{Andrews1999}. Then we see that
\begin{align}
\sum_{\gamma =1} ^{k}
\left(-1 \right)^{\gamma -1}
{1 \over \gamma} 
{k-1 \choose \gamma- 1}  
{n-k-1 \choose \gamma- 1} 
&=
{}_2F_1(-(k-1), -\{(n-k)-1\}; 2 ; -1)
\nonumber \\
&=
(1/2)^{-(k-1)} {}_2F_1(-(k-1), n-k+1; 2 ; 1/2)
\nonumber \\
&=
{1 \over k} (1/2)^{-(k-1)} P^{1,n-2k} _{k-1}(0).
\label{yukari1}
\end{align}
The first equality is given by the definition of the hypergeometric series. The second equality comes from the following relation:
\[
{}_2F_1(a, b; c ;z) = (1-z)^{-a} {}_2F_1(a, c-b; c ;z/(z-1)).
\]
The last equality follows from Eq. (\ref{konno-eqn:ken}). In a similar way, we have
\begin{align}
\sum_{\gamma =1} ^{k}
\left( -1 \right)^{\gamma -1}
{k-1 \choose \gamma- 1}  
{n-k-1 \choose \gamma- 1} 
= (1/2)^{-(k-1)} P^{0,n-2k} _{k-1}(0).
\label{yukari2}
\end{align}
By using Eqs. (\ref{yukari1}) and (\ref{yukari2}), we get
\begin{lem}
\label{sirowine}
For $1 \le l \le [n/2]$, 
\begin{align*}
p^{(H)}_n (l,n-l) 
&= \left( \frac{1}{\sqrt{2}} \right)^{n-2l+1} \> (-1)^{n-l} \left\{ P^{0,n-2l}_{l-1} (0) -  P^{1,n-2l}_{l-1} (0) \right\}, \\
q^{(H)}_n (l,n-l) 
&= \left( \frac{1}{\sqrt{2}} \right)^{n-2l+1} \> (-1)^{n-l} \> \left\{ \left( \frac{n-l}{l} \right) P^{1,n-2l}_{l-1} (0) -  P^{0,n-2l}_{l-1} (0) \right\}.
\end{align*}
\end{lem}
Finally we will consider the limit behavior of the characteristic function $E^{\omega}_n (e^{i \xi X_n/n})$ with $\xi \in \RM$ as $n \to \infty$. From Proposition {\rmfamily \ref{pro2}}, we have 
\begin{align}
E^{\omega}_n (e^{i \xi X_n/n}) 
&= E^{{\bf 0}}_n (e^{i \xi X_n/n}) 
+ \frac{\sin (\omega_0)}{2} \> \sum_{l=0}^n \> e^{i \xi (n-2l)/n} \left\{ p^{(H)}_n (l,n-l)^2 - q^{(H)}_n (l,n-l)^2 \right\}.
\label{akawine}
\end{align}
For the first term of the right-hand side of Eq. (\ref{akawine}) corresponding to the symmetric Hadamard walk, Theorem {\rmfamily \ref{thm1}} implies
\begin{align}
\lim_{n \to \infty} E^{{\bf 0}}_n (e^{i \xi X_n/n}) 
= \int_{-1/\sqrt{2}}^{1/\sqrt{2}} e^{i \xi x} {1 \over \pi (1 - x^2) \sqrt{1 - 2 x^2}} dx. 
\label{yyukari1}
\end{align}
Concerning the second term of the right-hand side of Eq. (\ref{akawine}), by a similar argument in \cite{Konno2002, Konno2005}, we see that 
\begin{align}
&\lim_{n \to \infty} \frac{1}{2} \> \sum_{l=0}^n \> e^{i \xi (n-2l)/n} \left\{ p^{(H)}_n (l,n-l)^2 - q^{(H)}_n (l,n-l)^2 \right\} \nonumber \\
&=\lim_{n \to \infty} 
\sum_{l=0}^n \> e^{i \xi (n-2l)/n} \left( \frac{1}{2} \right)^{n-2l+2} \left[ \left\{ \left(P^{1,n-2l}_{l-1} \right)^2 - 2 P^{1,n-2l}_{l-1} \> P^{0,n-2l}_{l-1} + \left( P^{0,n-2l}_{l-1} \right)^2 \right\} \right. \nonumber \\
&\qquad \qquad - \left. \left\{ \left( \frac{n-l}{l} \right) \left(P^{1,n-2l}_{l-1} \right)^2 - \frac{2(n-l)}{l} P^{1,n-2l}_{l-1} \> P^{0,n-2l}_{l-1} + \left( P^{0,n-2l}_{l-1} \right)^2 \right\} \right] \nonumber \\
&= \int_{(1-1/\sqrt{2})/2}^{(1+1/\sqrt{2})/2} \> e^{i \xi (1-2x)} \left( \frac{1}{2} \right) \> \frac{1}{\pi \sqrt{1-2(1-2x)^2}} \nonumber \\
&\qquad \qquad \times \left[ \left\{ 1 - \left( \frac{1-x}{x} \right)^2 \right\} \> \frac{2x}{1-x} - 2 \left(1 - \frac{1-x}{x} \right) \> \frac{1}{2(1-x)} \right] dx \nonumber \\
&= \int_{(1-1/\sqrt{2})/2}^{(1+1/\sqrt{2})/2} \> e^{i \xi (1-2x)} \frac{1}{2 \pi \sqrt{1-2(1-2x)^2}} \> \frac{2x-1}{x(1-x)} \> dx \nonumber \\
&= - \int_{-1/\sqrt{2}}^{1/\sqrt{2}} e^{i \xi x} 
\frac{x}{\pi (1 - x^2) \sqrt{1 - 2x^2}} \> dx, 
\label{yyukari2}
\end{align}
where $P^{i,n-2l} _{l-1} = P^{i,n-2l} _{l-1}(0) \> (i=0,1)$. The first equality comes from Lemma {\rmfamily \ref{sirowine}}. Combining Eqs. (\ref{akawine}) and (\ref{yyukari1}) with Eq. (\ref{yyukari2}) gives 
\begin{align*}
\lim_{n \to \infty} E^{\omega}_n (e^{i \xi X_n/n}) 
= \int_{-1/\sqrt{2}}^{1/\sqrt{2}} e^{i \xi x} \> \frac{1 - \sin (\omega_0) x}{\pi (1 - x^2) \sqrt{1 - 2 x^2}} \> dx,
\end{align*}
so the proof of Theorem {\rmfamily \ref{thm2}} is complete.



\begin{small}
\bibliographystyle{jplain}

\end{small}

\end{document}